\documentclass[conference,letter]{IEEEtran}

\usepackage[utf8]{inputenc} 
\usepackage{graphicx}
\usepackage[T1]{fontenc}
\usepackage{url}              
\usepackage{cite}             
\usepackage[cmex10]{amsmath}  
\usepackage{mleftright}       
\mleftright                   

\usepackage{graphicx}         
\usepackage{booktabs}         

\usepackage{balance}

\usepackage{mathpazo}
\usepackage{times}

\usepackage[T1]{fontenc}
\usepackage{amsmath,amssymb,amsfonts}
\usepackage{mathrsfs}
\usepackage{mathabx}
\usepackage{amsbsy}
\usepackage{graphicx}
\usepackage{graphics}
\usepackage{epstopdf}
\usepackage{algorithm}
\usepackage{algorithmic}
\usepackage{subfigure}
\usepackage{cite}
\usepackage{multirow}

\usepackage{caption}
\usepackage{tikz,pgfplots}
\usepackage{ctable}
\usepackage{setspace}
\usepackage{mathtools}

\newtheorem{theorem}{{Theorem}}
\newtheorem{lemma}[theorem]{{Lemma}}

\newcommand{\cB}{{\cal B}}
\newcommand{\cC}{{\cal C}} 

\newcommand{\cE}{{\cal E}}

\newcommand{\cH}{{\cal H}}
\newcommand{\cI}{{\cal I}}

\newcommand{\cR}{{\cal R}}

\DeclareMathAlphabet{\mathbfsl}{OT1}{ppl}{b}{it} 

\newcommand{\bC}{\mathbfsl{C}} 

\newcommand{\bE}{\mathbfsl{E}}

\newcommand{\bL}{\mathbfsl{L}}

\newcommand{\bR}{\mathbfsl{R}}

\newcommand{\bU}{\mathbfsl{U}}

\newcommand{\bX}{\mathbfsl{X}}

\newcommand{\ba}{\mathbfsl{a}} 
\newcommand{\bb}{\mathbfsl{b}} 
 
\newcommand{\bu}{\mathbfsl{u}} 
\newcommand{\bv}{\mathbfsl{v}}

\newcommand{\br}{\mathbfsl{r}}
\newcommand{\bl}{\mathbfsl{l}}

\newcommand*{\rom}[1]{\expandafter\romannumeral #1}

\makeatletter
\newcommand{\AlignFootnote}[1]{%
	\ifmeasuring@
	\else
	\iffirstchoice@
	\footnote{#1}%
	\fi
	\fi}
\makeatother

\newcommand{\be}[1]{\begin{equation}\label{#1}}
\newcommand{\ee}{\end{equation}} 

\newcommand{\lb}[1]{\label{#1}}

\renewcommand{\leq}{\leqslant}
 
\renewcommand{\geq}{\geqslant}

\renewcommand{\Bbb}{\mathbb}

\newcommand{\R}{{\Bbb R}}

\newcommand{\F}{{\Bbb F}}

\newcommand{\Lref}[1]{Lem\-ma\,\ref{#1}}
\newcommand{\Cref}[1]{Co\-ro\-lla\-ry\,\ref{#1}}

\newcommand{\Fq}{{{\Bbb F}}_{\!q}}

\newcommand{\bin}{\{0,1\}}

\newcommand{\deff}{\mbox{$\stackrel{\rm def}{=}$}}

\newcommand{\one}{{\mathbf 1}}

\newcommand{\norm}[1]{\left\lVert#1\right\rVert}

\usepackage{graphicx}  
\usepackage{float}
\usepackage[labelformat=simple,caption=false]{subfig}
\renewcommand{\baselinestretch}{1.0} 
\begin{document}
\IEEEoverridecommandlockouts
\title{{Matrix Completion over Finite Fields: \\Bounds and Belief Propagation Algorithms}
\thanks{This work was supported by the Department of Energy under grant DE-SC0022186.\\
Mahdi Soleymani is with Halıcıo\u{g}lu Data Science Institute, University of California San Diego. Qiang Liu is with the State Key Laboratory of Industrial Control Technology, Institute of Cyber-Systems and Control, Zhejiang University,
Hangzhou 310027, China. Hessam Mahdavifar, and Laura Balzano are with the Department of Electrical Engineering and Computer Science, University of Michigan Ann Arbor (e-mail: msoleymani@ucsd.edu, qiangliu\_421@zju.edu.cn, hessam@umich.edu, girasole@umich.edu).}} 


\author{%
  \IEEEauthorblockN{Mahdi Soleymani, Qiang Liu, Hessam Mahdavifar, and Laura Balzano}}





\maketitle

\begin{abstract}
We consider the low rank matrix completion problem over finite fields. This problem has been extensively studied in the domain of real/complex numbers, however, to the best of authors' knowledge, there exists merely one efficient  algorithm to tackle the problem in the binary field, due to Saunderson et al. \cite{saunderson2016simple}. In this paper, we improve upon the theoretical guarantees for the algorithm provided in \cite{saunderson2016simple}. Furthermore, we  formulate a new graphical model for the matrix completion problem over the finite field of size $q$, $\F_q$, and present a message passing (MP) based approach to solve this problem. The proposed algorithm is the first one for the considered matrix completion problem over finite fields of arbitrary size.  Our proposed method has a significantly lower computational complexity, reducing it from $O(n^{2r+3})$ in \cite{saunderson2016simple} down to $O(n^2)$ (where, the underlying matrix has dimension $n \times n$ and $r$ denotes its rank), while also improving the performance.

\end{abstract}

\section{Introduction}

The \emph{low rank matrix completion} problem aims at recovering the missing entries of a low-rank matrix $\bX$ by observing a \emph{small} fraction of its entries\cite{berry1999matrices}. This implies that many entries of $\bX$ are redundant and can be discarded for many large-scale scientific computations. This perspective has been used, for example, to speed up tasks in video processing by orders of magnitude \cite{he2012incremental} by processing only very small subsets of pixels from each frame. 
There exists an extensive literature studying the low rank matrix completion problem when the underlying matrix is over the field of real/complex numbers, see, e.g., \cite{candes2012exact}. Moreover, several polynomial-time algorithms exist, including optimization-based methods, that  provably recover the underlying real-valued matrices \cite{davenport2016overview}\cite{recht2010guaranteed}. However, such methods can not be applied to the case where the matrix under consideration is over the finite field $\F_q$.


Finite-field matrix completion is, by nature, a different problem compared to the real-field matrix completion, due to the fundamental differences in the underlying algebraic structures of finite fields and the infinite fields of real/complex numbers. The problem has several important applications in network coding \cite{birk1998informed, byrne2018index}, index coding \cite{esfahanizadeh2014matrix}, and decoding rank-metric codes under erasures \cite{martinez2022codes}. Several prior works \cite{saunderson2016simple, tan2011rank, ganian2018parameterized, vishwanath2010information} have considered rank metric and Hamming metric to establish sample complexity requirements and error bounds in finite-field matrix completion. Another closely related problem is Boolean matrix factorization\cite{miettinen2020recent, ravanbakhsh2016boolean} where the goal is to decompose $\bX$ into the Boolean multiplication of two matrices. 
The main distinction of our setup compared with this line of work is that the underlying constraints in our model are bilinear constraints over $\F_q$ whereas the constraints in \cite{miettinen2020recent, ravanbakhsh2016boolean} involve logical AND/OR operations on Boolean variables.



In this paper, we consider the low rank matrix completion over a finite field $\F_q$. 
This problem  has been proved to be $NP$-hard, including the case where the entries are over $\F_2$ \cite{peeters1996orthogonal}. A related earlier work \cite{saunderson2016simple} showed that the matrix completion problem over $\F_2$ can be solved  with high probability and  proposed a linear programming-based algorithm to tackle this problem. Moreover, \cite{ganian2018parameterized} proved this problem is fixed-parameter tractable over prime fields. 
However,  efficient algorithms for  matrix completion problem over a finite field with arbitrary size do not exist.

The contribution of this paper is twofold. In the first part, we improve upon the theoretical recovery guarantees of the algorithm proposed in \cite{saunderson2016simple} for matrix completion over $\F_2$. Specifically, we improve the threshold for the probability of observation of the entries for the algorithm in \cite{saunderson2016simple} to successfully recover $\bX$ by making a connection between the probability of unsuccessful recovery  in \cite{saunderson2016simple} and the error probability of maximum likelihood (ML) decoder for random binary linear codes over the binary erasure channel (BEC). In the second part, we characterize a new graphical model framework for matrix completion over finite fields and utilize a variant of message-passing algorithm to arrive at a solution. Specifically, we run the \emph{sum-product} algorithm for several rounds repeatedly for a fixed number of iterations, where at the end of each round the value of one randomly picked variable node is fixed. This procedure is continued until we either converge to a solution or the maximum number of rounds is reached. A similar idea is  known in the literature as the  belief propagation guided decimation (BPGD) approach and has been utilized in the context of the $k$-satisfiability ($k$-SAT) problem \cite{montanari2007solving, mezard2002analytic}. However, the bilinear constraints in our proposed factor graph  involve addition and multiplication operations over $\F_q$ whereas the constraints in the $k$-SAT problem merely involve logical OR operations.  The main distinction between our decimation procedure and the existing ones is that we do not check for any possible contradictions between the variables fixed so far at each intermediate round as it is done in \cite{montanari2007solving, mezard2002analytic} in a subroutine called warning propagation (WP). It turns out that this relaxation does not alter the performance of the algorithm.  Our  empirical studies demonstrate the superiority of our algorithm over $\F_2$  in terms of both the performance and the computational complexity compared to \cite{saunderson2016simple}.

\section{Notations}
For a $p \in [0,1]$ and a finite set $A$, $S \sim \cB(A,p)$ is the random subset of $A$ obtained by choosing each element of $A$ independently with probability $p$. 
An $(n,k)$ linear code $\cC$ is a $k$-dimensional subspace of the $\F_2^n$ vector space. The minimum Hamming distance of $\cC$ is denoted by $d_{\min}(\cC)$, and the dimension of $\cC$ is also denoted by $\dim \cC$.
The $\otimes$ notation denotes an operation similar to inner product over finite field vectors, that is, for $\ba, \bb \in \F_q^{n}$ we have $\ba \otimes \bb \deff \sum_{i=1}^{i=n} \ba(i) \times \bb(i)$, where $\times$ denotes the finite field multiplication. The function nnz$(\cdot)$ returns the number of non-zero elements of its vector/matrix input.


\section{Matrix Completion over Finite Fields}

In this section, we consider the problem of  matrix completion over a finite field where we aim at recovering a low-rank matrix based on a partial observation of its entries. Let $\bX \in \Fq^{m \times n}$ denote a matrix of rank $r$. Each entry of $\bX$ is revealed by probability $p$, independent of all other entries. Specifically, $P_\Omega(\bX)$ is observed where $\Omega \sim \cB(p, [m] \times [n])$. The goal is to fully recover $\bX$ using this partial observation. This problem is known to be NP-hard \cite{harvey2006complexity} and, hence, the refined goal is to 
obtain algorithms that recover $\bX$ with high probability and with polynomial time complexity. This problem has been studied in \cite{saunderson2016simple} over $\F_2$ and a simple algorithm is provided for the matrix completion over $\F_2$ with $O(n^{2r+3})$ complexity for $m=n$, which is polynomial in $n$ when $r$ is fixed. In this section, we improve the theoretical guarantees on the algorithm provided in \cite{saunderson2016simple}  modifying their analysis on the probability of recovery error. 

 The following lemma is used repeatedly in the analysis provided in \cite{saunderson2016simple}.
\begin{lemma}\cite[Lemma\,1]{saunderson2016simple}\lb{Saunderson}
Let $\cC \in \F_2^n$ be a binary linear code. If $S \sim \cB([n],p)$ then 
\be{Saunderson_bound}
{\small \Pr [P_s(x)=0 \text{ for some } x \in \cC \backslash \{0\}] \leq 2^{\dim \cC}e^{-pd_{\min}(\cC)}.}
\ee
\end{lemma}
\begin{figure*}[htb!]
	\begin{minipage}{0.32\textwidth} \vspace{-.4cm}
		\centering
%

\begin{tikzpicture}

\definecolor{mycolor1}{rgb}{0.15,0.15,0.15}
\definecolor{mycolor2}{rgb}{0,0,1}
\definecolor{mycolor3}{rgb}{1,0,0}
\definecolor{mycolor4}{rgb}{0,0.45,0.74}
\definecolor{mycolor5}{rgb}{0.64,0.08,0.18}
\definecolor{mycolor6}{rgb}{0.2,0.6,0.2}
\definecolor{mycolor7}{rgb}{1,0 ,1}

\begin{axis}[
x label style={at={(axis description cs:0.47,0)},anchor=north},
y label style={at={(axis description cs:0.05,0.5)},rotate=0,anchor=south},
xlabel={\small $n$},
ylabel={ \small Lower bound on $p$},
scale only axis,
every outer x axis line/.append style={mycolor1},
every x tick label/.append style={font=\color{mycolor1}},
every outer y axis line/.append style={mycolor1},
every y tick label/.append style={font=\color{mycolor1}},
width=1.5in,
height=1in,
xmin=20, xmax=50,
ymin=0.6, ymax=1,
axis on top,
legend entries={{\small $\tilde{p}$: Our lower bound }, {\small $p'$: Lower bound in \cite{saunderson2016simple}}. },
legend style={ nodes={scale=.9, transform shape}, legend columns=1,at={(0,1.25)},anchor=west},
legend cell align=right,
mark options={solid,scale=1.3}
]
\addplot [
color=blue,
solid,
line width=1.0pt,
]
coordinates{
	(20,0.839702)
	(21,0.827427)
	(22,0.815598)
	(23,0.804248)
	(24,0.793392)
	(25,0.783026)
	(26,0.773129)
	(27,0.763667)
	(28,0.754594)
	(29,0.745855)
	(30,0.737391)
	(31,0.72914)
	(32,0.721387)
	(33,0.714054)
	(34,0.706785)
	(35,0.699516)
	(36,0.692967)
	(37,0.686463)
	(38,0.679784)
	(39,0.67396)
	(40,0.667899)
	(41,0.662074)
	(42,0.656574)
	(43,0.650844)
	(44,0.645821)
	(45,0.640278)
	(46,0.635626)
	(47,0.630352)
	(48,0.625949)
	(49,0.621012)
	(50,0.616729)
	
};

\addplot [
color=red,
solid,
line width=1.0pt,
]
coordinates{
	(30,1)
	(31,0.984409)
	(32,0.96854)
	(33,0.953663)
	(34,0.93923)
	(35,0.925806)
	(36,0.912742)
	(37,0.900088)
	(38,0.88833)
	(39,0.876983)
	(40,0.86604)
	(41,0.855499)
	(42,0.84535)
	(43,0.835572)
	(44,0.826136)
	(45,0.817009)
	(46,0.808148)
	(47,0.799509)
	(48,0.791377)
	(49,0.783583)
	(50,0.775913)
	
};

\end{axis}

\end{tikzpicture}\vspace{-.9cm}
		\caption{ \small Comparison between $\tilde{p}$ and $p'$. Other parameters are $k=2$ and $\theta=0.1$.}  \label{p_1}
	\end{minipage}\hfill
 	\begin{minipage}{0.32\textwidth} \vspace{-.5cm}
		\centering
%

\begin{tikzpicture}

\definecolor{mycolor1}{rgb}{0.15,0.15,0.15}
\definecolor{mycolor2}{rgb}{0,0,1}
\definecolor{mycolor3}{rgb}{1,0,0}
\definecolor{mycolor4}{rgb}{0,0.45,0.74}
\definecolor{mycolor5}{rgb}{0.64,0.08,0.18}
\definecolor{mycolor6}{rgb}{0.2,0.6,0.2}
\definecolor{mycolor7}{rgb}{1,0 ,1}

\begin{axis}[
x label style={at={(axis description cs:0.47,0)},anchor=north},
y label style={at={(axis description cs:0.05,0.5)},rotate=0,anchor=south},
xlabel={\small $n$},
ylabel={ \small Lower bound on $p$},
scale only axis,
every outer x axis line/.append style={mycolor1},
every x tick label/.append style={font=\color{mycolor1}},
every outer y axis line/.append style={mycolor1},
every y tick label/.append style={font=\color{mycolor1}},
width=1.5 in,
height=1in,
xmin=20, xmax=50,
ymin=0.8, ymax=1,
axis on top,
legend entries={{\small $\tilde{p}$: Our lower bound }, { \small $p'$: Lower bound in \cite{saunderson2016simple}}. },
legend style={ nodes={scale=.9, transform shape}, legend columns=1,at={(-.01,1.25)},anchor=west},
legend cell align=right,
mark options={solid,scale=1.3}
]
\addplot [
color=blue,
solid,
line width=1.0pt,
]
coordinates{
	(20,0.972686)
	(21,0.965913)
	(22,0.958414)
	(23,0.950838)
	(24,0.943893)
	(25,0.936674)
	(26,0.929225)
	(27,0.922605)
	(28,0.916116)
	(29,0.909223)
	(30,0.903292)
	(31,0.897236)
	(32,0.890968)
	(33,0.885722)
	(34,0.879604)
	(35,0.874767)
	(36,0.869097)
	(37,0.864403)
	(38,0.859038)
	(39,0.854609)
	(40,0.849386)
	(41,0.845319)
	(42,0.840131)
	(43,0.836432)
	(44,0.831841)
	(45,0.827831)
	(46,0.823864)
	(47,0.819395)
	(48,0.816042)
	(49,0.811882)
	(50,0.808236)
	
};

\addplot [
color=red,
solid,
line width=1.0pt,
]
coordinates{
	(46,1)
	(47,0.987626)
	(48,0.980193)
	(49,0.973883)
	(50,0.967237)
	
};

\end{axis}

\end{tikzpicture}\vspace{-.9cm}
		\caption{  \small Comparison between $\tilde{p}$ and $p'$. Other paramers are $k=3$ and $\theta=0.1$.}\label{p_2}
	\end{minipage}\hfill
  	\begin{minipage}{0.32\textwidth}
		\centering\vspace{1cm}
		\includegraphics[height=2.7cm]{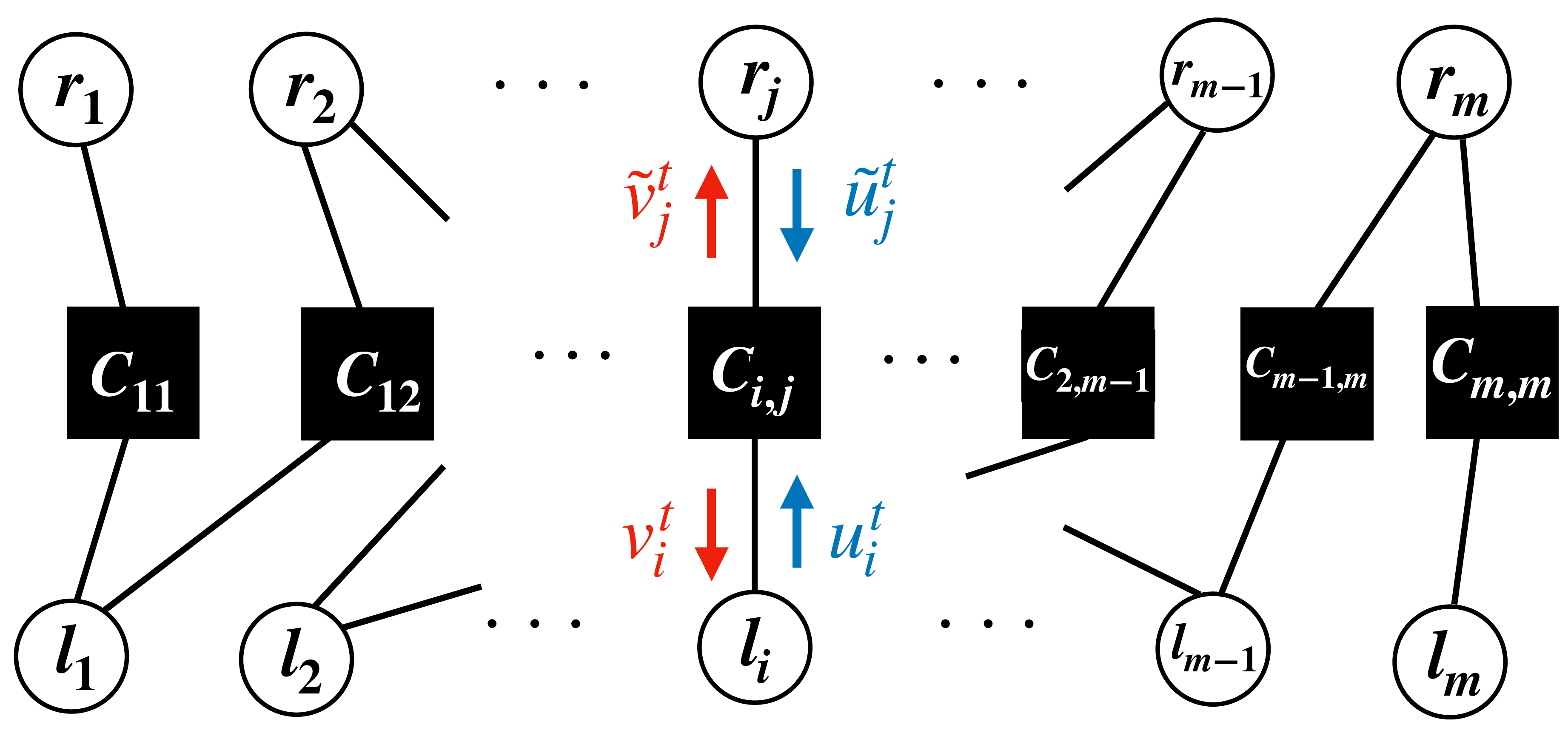} \vspace{.6cm}
		\caption{ \small Illustration of the proposed factor graph for matrix completion problem over $\F_{q}$.
  The variable nodes in the top and the bottom layers correspond to the rows of $\bL$ and the columns of $\bR$, respectively.  
  }\label{mp}
	\end{minipage}\hfill
 \vspace{-5mm}
\end{figure*}
		
Let $P_e(\epsilon,k,n)$ denote the probability of error under maximum likelihood (ML) decoding  over a binary erasure channel (BEC) for an $(n,k)$ binary linear code. Then, one can observe that 
\begin{equation*}
 \Pr [P_s(x)=0 \text{ for an } x \in \cC \backslash \{0\}] = P_e(\epsilon=1-p, n,\dim \cC). 
\end{equation*}
Note that the upper bound provided in \Lref{Saunderson} only utilizes the minimum distance of the code and is not tight.
In light of this observation, we  replace this upper bound with a tighter one for the probability of error of a random binary linear code over BEC. The ensemble of random binary linear codes, denoted by $\cR(n,k)$, is obtained by picking each entry of the $k \times n$ generator matrix independently and uniformly at random. 
The average probability of error of a code picked randomly from $\cR(n,k)$ over BEC$(\epsilon)$ under ML decoding is equal to (see, e.g., \cite{soleymani2021coded, jamali2019coded}):
\begin{equation}\label{ashikhmin}
 {P_e(\epsilon, k,n)= \sum_{e=0}^{n} {n \choose e} \epsilon^e (1-\epsilon)^{n-e} [1-\prod_{i=n-e-k+1}^{n-e} (1-\frac{1}{2^i})]}.
\end{equation}
\noindent
By using the result on the minimum distance of the random binary linear codes \cite{barg2002random}, we can compare the upper bound provided in \eqref{Saunderson_bound} with $P_e(\epsilon,k,n)$ characterized in \eqref{ashikhmin} in Figure\,\ref{logpe_comp} in the Appendix.
The numerical evaluations, as illustrated in  Figure\,\ref{logpe_comp} in the Appendix shows a significant gap between the guarantees on $P_e(\epsilon,k,n)$ characterized in this paper compared to that derived in \cite{saunderson2016simple}. This motivates us to determine how this alters the theoretical guarantees on recovering $\bX$ in the matrix completion problem. For the case where the entries $\bX $ are drawn uniformly at random,  it is shown in \cite{saunderson2016simple} that the proposed algorithm can recover $\bX$ with probability at least $1-3\theta$ when $p>p'$ for some  $p'$ that is a function of system parameter and $\theta$. By following the exact steps in the proofs in \cite{saunderson2016simple} and replacing \eqref{Saunderson_bound} by \eqref{ashikhmin}, one can show that the matrix $\bX$ can be recovered  for  $p>\tilde{p}$, where $\tilde{p}>p'$. The parameters $p'$ and $\tilde{p}$ are the solutions to 
\be{eq}
n^{r+1}P_e(1-p^{r+1},n,r)=\theta,
\ee
when \eqref{Saunderson_bound} and \eqref{ashikhmin} are used, respectively (see Appendix\,\ref{derivation} for details). In Figures\,\ref{p_1} and \ref{p_2}, we compare the numerical evaluations of $p'$ and $\tilde{p}$ for a certain set of parameters. Note that there exist regions that the analysis in \cite{saunderson2016simple} does not provide a non-trivial bound on $p$. 


\section{A massage passing algorithm for Matrix Factorization over $\F_{q}$}
\subsection{Graphical model}
Let $[n]=\{1,2,...,n\}$. Let $\bX \in \F_{q}^{m \times n}$ denote a rank-$r$ matrix whose entries are observed according to the observation matrix $\Omega \in {\{0,1\}}^{m \times n}$, i.e., $\bX_{ij}$ is revealed if and only if $\Omega_{ij}=1$, for all $i \in [m]$ and $j \in [n]$. Note also that  $\bX=\bL\bR$ for some  $\bL \in \F_{q}^{m \times r}$ and $\bR \in \F_{q}^{m \times r}$ since the rank of $\bX$ is $r$. This decomposition is not  unique, i.e., there  exist several distinct pairs of $\bL$ and $\bR$ such that $\bX=\bL\bR$.
Our goal in this section is to find one such pair. 
Next, we propose a message passing (MP) algorithm to solve this problem.

Figure\,\ref{mp} illustrates the factor graph considered for the matrix factorization problem. Each variable node $l_i$ corresponds to a row in $\bL$ for all $i \in [m]$. Similarly, each variable node $r_j$ corresponds to a column in $\bR$ for all $j \in [n]$. If $\Omega_{ij}=1$, i.e., the $(i,j)$ entry of $\bX$ is observed, there exist a factor node $C_{i,j}$ with two variable nodes connected to it; $\bl_i$ and $\br_j$. Note that each factor graph is connected to exactly two variable nodes, where each variable node could be connected to several factor nodes. We regard this problem as a constraint satisfaction problem and leverage the well-known message passing algorithm that attempts to find a solution that satisfies all constraints. Specifically, we use the so-called \emph{Sum-Product} MP algorithm \cite{kschischang2001factor} to compute the marginals of the variable nodes in the factor graph.  

 Let $\mu_i$ for $i \in [m]$ and $\nu_j$ for $j \in [n]$ denote the marginal distributions of $\boldsymbol{l}_i$ and $\boldsymbol{r}_j$, respectively. Note  that there are  $q^{r}$ possibilities for each variable node, i.e., $\mu_i$'s and $\nu_j$'s are tuples with $q^{ r}$ elements.  
Let $\bv_i^t$ and $\bv_j^t$ denote the message sent from $C_{i,j}$ to $\boldsymbol{l}_i$  and $\boldsymbol{r}_j$ at iteration $t$, respectively. Let also $\bu_i^t$ and $\bu_j^t$ denote the message sent from $\boldsymbol{l}_i$  and $\boldsymbol{r}_j$  to $C_{i,j}$ at iteration $t$, respectively. For all $i \in [m]$, let $\cC_i^L$ denote the set of indices $j$ such that $\Omega_{ij}=1$, i.e., the indices corresponding to the variable nodes $\br_j$ that are connected to  the node $\bl_i$ through a factor node, namely $C_{i,j}$. Similarly, let $\cC_j^R$ denote the set of all indices $i$ such that $\Omega_{ij}=1$. Recall that all the messages are a probability distribution over a set of $q^{ r}$ variables. The $i$-th entry of a vector $\bu$ is denoted by $\bu(i)$. The Hadamard product between two vectors $\bl\deff \bu \circ \bv$ is a vector consisting of the element-wise product  of $\bu$ and $\bv$, i.e., $\bl(i)=\bu(i)\bv(i)$. 

The constraints in the factor graph illustrated in Figure\,\ref{mp} are 
\be{constraints}
\bl_i \otimes \br_j = \bX_{ij}\quad \quad \forall \quad i, j \quad \text{s.t.} \quad \Omega_{i,j}=1.
\ee
Note that the constraints are bilinear over $\F_q$. Similar factor graphs with bilinear constrains over $\R$ are considered in  \cite{parker2014bilinear, parker2014bilinear2}  and  approximate message passing algorithms are specialized  for matrix completion, robust PCA, etc. 
\subsection{Message passing algorithm}
Let $\boldsymbol{\alpha} (\cdot): [q^r] \xrightarrow[]{} \F_q^r$ denote a one-to-one mapping from $[q^r]$ to a the set of vectors of length $r$ over $\F_q$. 
Then, the SP update equations for the message passing algorithm over the factor graph illustrated in Figure\,\ref{mp} under the constraints characterized in \eqref{constraints} can be written as 
\begin{align}
\bu_i^t&\cong\prod_{\Omega_{i,j}=1,  j\in[n] } \bv_i^{t-1} \lb{a1}\\
\tilde{\bu}_j^t&\cong \prod_{\Omega_{i,j}=1,  j\in[n] } \bv_j^{t-1} \lb{a2}\\
\small \bv_i^t(l)\cong \sum_{k\in [q^{ r}]} \bu_j^{t-1}(k) & \cI(\boldsymbol{\alpha} (k)\otimes \boldsymbol{\alpha}(l) =\bX_{ij}) \lb{a3}\\
\tilde{\bv}_j^t(l) \cong \sum_{k\in [q^{ r}]} \bu_i^{t-1}(k) & \cI(\boldsymbol{\alpha} (k)\otimes \boldsymbol{\alpha}(l) =\bX_{ij}), \lb{a4}
\end{align}
where $\cong$ denotes equality up to a normalization constant, and $\cI$ denotes the indicator function. \
Utilizing the SP algorithm to find the marginal distributions of the variable nodes over a factor graph is not new, neither the update equations in \eqref{a1}-\eqref{a4}. However, we can further simplify equations  \eqref{a1}-\eqref{a4} and write the update equations for $\bu_i^t$ and $\tilde{\bu}_j^t$ in terms of $\tilde{\bu}_j^{t-1}$ and $\bu_i^{t-1}$. Note that we only require the final values of $\tilde{\bu}_j^t$ and ${\bu}_j^t$ after running the SP algorithm. 
Therefore, one can re-write the update equations for $\tilde{\bu}_j^t$ and ${\bu}_j^t$ as multiplying a certain matrix, as characterized below, by $\tilde{\bu}_j^{t-1}$ and ${\bu}_j^{t-1}$, respectively. This will significantly reduce the computational complexity of the implementation.\

Let  $\bC\in \{0,1\}^{q^{r} \times q^{ r}} \times q$ denote a matrix whose $(i,j,\beta(\bX_{i,j}))$ entry is equal to $1$ if $\boldsymbol{\alpha} (k)\otimes \boldsymbol{\alpha}(l) =\bX_{ij}$, and $0$ otherwise. By substituting \eqref{a3} and \eqref{a4} into \eqref{a1} and \eqref{a2}, respectively, one can write 
\begin{align}
\bu_i^t \cong  \prod_{\gamma \in \F_q} \prod_{j \in \cC_i^L, \bX_{ij}=\gamma} \left(\bC[:, :, \beta(\gamma)] \tilde{\bu}_j^{t-1}\right)  , \ \forall i \in [m] \label{u_i_update}\\
\tilde{\bu}_j^t \cong  \prod_{\gamma \in \F_q} \prod_{i \in \cC_j^R, \bX_{ij}=\gamma} \left(\bC[:,:,\beta(\gamma)] {\bu}_i^{t-1}\right) ,  \ \forall j \in [n], \label{u_j_update}
\end{align}
where all  the vector-vector multiplications are Hadamard multiplication, and, $\beta(\cdot): \F_q \xrightarrow[]{} [q]$ is a one-to-one mapping from $\F_q$ to $[q]$. Recall that all operations are performed over $\R$ as the belief vectors corresponding to the variable nodes are real-valued. 

It is well-known that if the underlying factor graph depicted in Figure\,\ref{mp}  is loop-free, then the SP algorithm is nothing but leveraging the distributive law that converges to the true marginals in a finite number of iterations. 
However, the SP algorithm is also often utilized for loopy factor graphs to approximate the solution. We consider a maximum number of iterations for SP and stop updating the beliefs if the number of iterations reaches this maximum, and the algorithm has not converged yet.

By updating the belief vectors according to the update equations in \eqref{a1}-\eqref{a4}, some of the multiplications could be computed several times when updating different beliefs. In order to avoid the unnecessary computation overhead, one can bring together all the belief vectors over $\bl_i$'s and $\br_j$'s into two matrices and multiply them to the constraint matrices once. The resulting matrices' columns can then be used to update the belief vectors according to \eqref{a1}-\eqref{a4}. Specifically, let $\bU_l^t$ and $\bU_r^t$ denote matrices whose columns consist of the beliefs over the variable nodes $\bl_i$'s and $\br_j$'s at iteration $t$, respectively, i.e.,  
\be{U}\bU^t \ \deff\
\begin{bmatrix}
	\bu_1^t &| & \bu_2^t &|& \cdots &|& \bu_m^t \\
\end{bmatrix},
\ee 
and,
\be{U_tilde}
\tilde{\bU}^t \ \deff\
\begin{bmatrix}
	\tilde{\bu}_1^t &| & \tilde{\bu}_2^t &|& \cdots &|& \tilde{\bu}_n^t 
\end{bmatrix},
\ee
and are referred to as \emph{belief matrices}. Then, one can implement the SP algorithm as provided in Algorithm\,\ref{sp-alg} to obtain an approximation of the marginal distributions of the variable nodes. This modification reduces the computational complexity of updating beliefs according to \eqref{u_i_update} and \eqref{u_j_update} by removing the  redundant matrix-vector multiplications.


\begin{figure*}[t]
\centering
\subfigure[Algorithm\,\ref{fovat}]{
\begin{minipage}[t]{0.3\linewidth}
\centering
\includegraphics[width=2in]{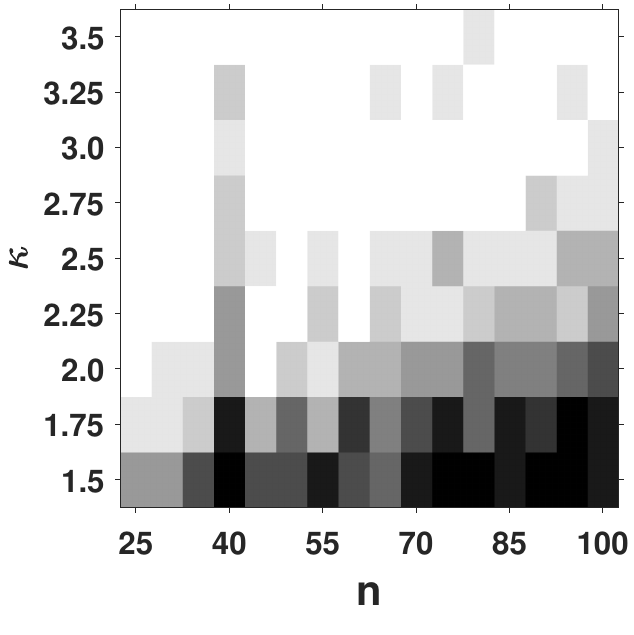}
\end{minipage}%
}%
\subfigure[Result in \cite{saunderson2016simple}]{
\begin{minipage}[t]{0.3\linewidth}
\centering
\includegraphics[width=2in]{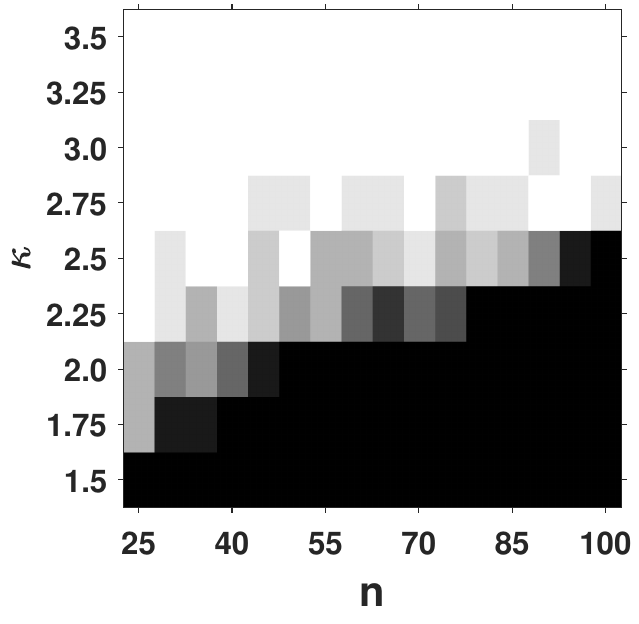}
\end{minipage}%
}%
\subfigure[Algorithm\,\ref{fovat}]{
\begin{minipage}[t]{0.3\linewidth}
\centering
\includegraphics[width=2in]{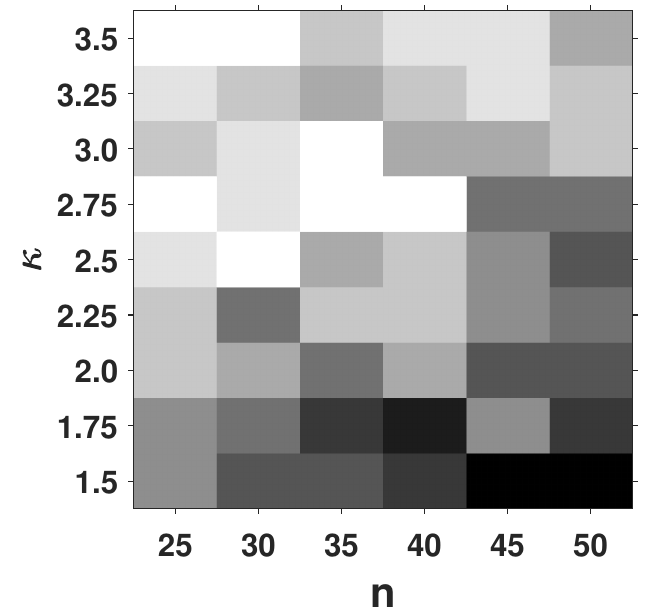}
\end{minipage}
}%
\centering
\caption{\small Plots (a) and (b) illustrate the performance of Algorithm\,\ref{fovat} versus the one in \cite{saunderson2016simple}. The matrix to be completed is $n\times n$ of rank $5$, entries are over $\F_2$ and revealed independently with probability $p = \kappa r(2n-r)/n^2$.
The grayscale intensity indicates the proportion of successful recoveries. The  corresponding pixel is black if the method failed on all attempts and white if it succeeded on all attempts. The number of trail is $10$ for each pixel.  
Plot (c) illustrates the performance of Algorithm\,\ref{fovat} over $\F_{2^3}$ for $r=3$ with $10$ trials for each pixel.
}
\label{F21}\vspace{-6mm}
\end{figure*}
\begin{algorithm}[!t]
	\textbf{Input:}  $\bX_\Omega \in \{0,1,\cE\}^{m \times n}$, $\Omega \in \bin^{m \times n}$, $r$, $t_{\max}$, $\epsilon_{\min}$.
 
	\textbf{Output:} Marginal beliefs $\bu_i $ and $\tilde{\bu}_j$ on $\bl_i$'s and $\br_j$'s. 
	
	\textbf{Initialization}: Set $\bU^0=\frac{1}{q^{r}} \one_{q^{r} \times m }$ and $\tilde{\bU}^0=\frac{1}{q^{r}} \one_{q^{r} \times n }$. Set $t=1$ and $\epsilon=\infty$. 
	
	\textbf{While} ($t \leq t_{\max}$) $\land$   ($ \epsilon_{\min} \leq \epsilon $) : 
	
	\quad \begin{align}
	\bE^t[:,:,\beta(\gamma)]=\bC[:,:,\beta(\gamma)] \bU^{t-1},\forall \gamma \in \F_q
	\\ \tilde{\bE}^t[:,:,\beta(\gamma)]=\bC[:,:,\beta(\gamma)]  \tilde{\bU}^{t-1},\forall \gamma \in \F_q
	\end{align}
	
	\quad \textbf{For} $i \in [m]$ and $j \in [n]$:
	\begin{align}
	\bu_i^t \cong \prod_{\gamma \in \F_q} \prod_{l \in \cC_i^L, \bX_{il}=\gamma} \tilde{\bE}_0[:,l,\beta(\gamma)] ,\\
	\tilde{\bu}_j^t \cong \prod_{\gamma \in \F_q} \prod_{l \in \cC_j^R, \bX_{lj}=\gamma} \bE_0[:,l,\beta(\gamma)] .
	\end{align}
	\quad \textbf{end;}
	
	\quad Construct $\bU^t$ and $\tilde{\bU}^t$ according to \eqref{U} and \eqref{U_tilde}.
	
	\quad $\epsilon=\min(\epsilon, \norm{\bU^t-\bU^{t-1}}+\norm{\tilde{\bU}^t-\tilde{\bU}^{t-1}})$.
	
	\quad $t=t+1.$
	
	\textbf{end;}
	
	\textbf{Return}: $\bU^t$ and $\tilde{\bU}^t$.
	\caption{SP algorithm for matrix completion over $\F_{q}$.}
	\label{sp-alg}
\end{algorithm}
The output of Algorithm\,\ref{sp-alg} does not uniquely determine a solution, neither it captures the correlations between different variable nodes. It merely provides the marginal distributions over $\bl_i$'s and $\br_j$'s, if it converges. Recall that the LR-factorization is not unique. Therefore the marginal beliefs over the variable nodes capture the likelihood of a certain configuration of variables over the set of all such LR-factorizations consistent with the observation. It is worth noting that, unlike running SP algorithm for the $k$-SAT problem, the SP algorithm over the factor graph depicted in Figure\,\ref{mp}  often converges to a solution very fast, but to a \emph{trivial} one. Specifically, \emph{almost} all the belief vectors converge to a uniform distribution that includes  all possible assignments (except for zero vectors) for $\bl_i$'s and $\br_j$'s. Roughly speaking, that means running the SP algorithm for a single round does not significantly \emph{reduce} the size of the search space of the variable nodes, let alone determining them. This motivates us to utilize a decimation procedure that guides us to a solution to the problem which is discussed next.

\subsection{Belief propagation-guided decimation algorithm}

In order to determine a solution, one can attempt to fix some of the variable nodes based on the associated beliefs returned by Algorithm\,\ref{sp-alg}, e.g., by sampling from the corresponding distribution. Since marginal beliefs capture no information about the statistical dependencies between the variable nodes, one might end up with an empty set of feasible solutions after fixing a few variable nodes according to their marginal beliefs. The challenge is that fixing one of the variable nodes alters the beliefs over other nodes through their statistical dependencies, rendering the marginals over other variable nodes incorrect after one variable is fixed by sampling. In order to overcome this problem, we propose utilizing a decimation  algorithm that fixes a random variable node at a time and runs the Algorithm\,\ref{sp-alg} again with the modified belief as its initial condition. This procedure is repeated for several rounds until all beliefs converge to \emph{single-entry} vectors, i.e., the vectors with only one non-zero element, or the maximum number of rounds $b_{\max}$ is reached. In the end, the value of the variable nodes is determined by sampling from the final beliefs.   Algorithm\, \ref{fovat} describes the steps of this decimation algorithm in detail.   By fixing a belief vector, e.g., a column of the belief matrix, we intend to replace it with a one-entry vector having a $1$ at a coordinate that is determined by random sampling according to the original vector. In other words, by fixing a belief vector, it collapses to a deterministic distribution with only one non-zero entry (which is equal to $1$). Algorithm\,\ref{fovat} runs in $O(mn)$ time complexity, provided that $t_{\max}$ and $b_{\max}$ are some constants.

\begin{algorithm}[h]
	\textbf{Input:}  $\bX_\Omega \in \{0,1,\cE\}^{m \times n}$, $\Omega \in \bin^{m \times n}$, $r$, $t_{\max}$, $\epsilon_{\min}$, $b_{\max}$.\\
	\textbf{Output:} Completed matrix $\tilde{\bX}$.
	
	\textbf{Initialization}: Set $\bU^i=\frac{1}{q^{r} }\one_{q^{ r} \times m }$ and $\tilde{\bU}^i=\frac{1}{q^{ r}} \one_{q^{ r} \times n }$, $b=1$, stop=$False$.
	
	\textbf{While}   $(b \leq b_{\max})\land (\text{converged}==False)$ :
	
	\quad \begin{itemize}
		
		\item 
		Run Algorithm\,\ref{sp-alg} with $\bU^i$ and $\tilde{\bU}^i$ as initial beliefs. 
		
		\item
		Denote the returned beliefs by  $\bU^o$ and $\tilde{\bU}^o$. 
		
		\item  Update $\bU^o$ by fixing all the columns  that were fixed during 
		previous steps.
		
		\item Update  $\bU^o$ by choosing a new column of it  that has not been fixed 
		before at random and fix it by sampling.
		
		\item \textbf{If} (nnz$(\bU^o)\leq m$) $\lor$  (nnz$(\tilde{\bU}^o)\leq m$): stop= $True$.
		
		\item Set $\bU^i=\bU^o$ and $\tilde{\bU}^i=\tilde{\bU}^o$, and, $b=b+1.$ 
		
	\end{itemize}
	
	\textbf{end}
	
	Construct $\bL$ and $\bR$ by sampling from the columns of  $\bU^o$ and $\tilde{\bU}^o$, respectively. 
	
	\textbf{Return}: $\tilde{\bX}=\bL \bR $. 
	
	\caption{Belief Propagation-Guided Decimation Algorithm for Matrix Completion over $\F_q$.}
	\label{fovat}
\end{algorithm}
\section{Experiments}
In this section, we compare the performance of Algorithm\,\ref{fovat} with that of the algorithm provided in \cite{saunderson2016simple} for matrix completion over $\F_2$. Furthermore, we  demonstrate the performance of our algorithm over $\F_{2^3}$ as a proof-of-concept example of a finite field with size larger than $2$.   

Figures\,\ref{F21} (a) and (b)  illustrate the comparison between the performance of our algorithm and that of the one provided in  \cite{saunderson2016simple}. 
The phase-transition plots illustrate the success of the two methods for completing  random $n \times n$ binary matrices of rank $r$ provided that the entries are revealed independently with probability $p = \kappa r(2n-r)/n^2$. The random matrix $\bX$ is generated by picking its left and right factor at random, that is, the entries of $\bL$ and $\bR$ are independently drawn from a uniform distribution over $\F_2$.  The pixel corresponding to $(r, \kappa)$ is black if the method fails on all attempts and white if it succeeds on all attempts. The results suggest that our algorithm  performs better comparing to the linear programming based approach in \cite{saunderson2016simple}. One can observe that when $p$ is relatively \emph{small}, the linear programming based approach is unable to recover the matrix in all trials while our algorithm recovers the matrix with high probability. Note that the computational complexity of Algorithm\,\ref{fovat} is $O(n^2)$ while the complexity of the algorithm in \cite[Theorem\,2]{saunderson2016simple} is $O(n^{2r+3})$ for completing an $n \times n$ matrix.
Figure\,\ref{F21} (c) demonstrates the performance of our algorithm over the extension field $\F_{2^3}$.  It shows that Algorithm\,\ref{fovat} recovers the matrix with relatively \emph{high} probability for a certain range of parameters of the problem. It is worth mentioning that this algorithm is the first efficient algorithm for matrix completion problem over a finite field with a general size, thereby establishing a benchmark for the performance of future algorithms. 


\clearpage 

\bibliographystyle{IEEEtran}
\bibliography{ref}
\newpage 
\appendices

\section{Derivation of \eqref{eq} as described in \cite{saunderson2016simple}}\label{derivation}
The meta algorithms proposed in \cite{saunderson2016simple} for matrix completion are all of the following form for different choices of $\cH_1$ and $\cH_2$.

1) Construct $\mathcal{H}_1 \subseteq 2^{\left[n_1\right]}$ and $\mathcal{H}_2 \subseteq 2^{\left[n_2\right]}$

2) For $i=1,2$ construct $U_i \in \mathbb{F}_2^{n_i \times k_i}$ with columns that are a basis for span $\left\{e_{S_i}: S_i \in \mathcal{H}_i\right\}^{\perp}$.

3) Return $U_1 \tilde{X} U_2^T$ for all $\tilde{X} \in \mathbb{F}_2^{k_1 \times k_2}$ satisfying
$$
P_{\Omega}\left(U_1 \tilde{X} U_2^T\right)=P_{\Omega}\left(X^{\star}\right)
$$
Let $\mathcal{H}_i \subseteq 2^{\left[n_i\right]}$ (for $i=1,2$ ) and let $P_{\Omega}\left(X^{\star}\right) \in$ $\mathbb{F}_2^{|\Omega|}$. We say that $\mathcal{H}_1$ and $\mathcal{H}_2$ are consistent with $P_{\Omega}\left(X^{\star}\right)$ if
\begin{equation}\label{consis}
\begin{aligned}
& \left\{X \in \mathbb{F}_2^{n_1 \times n_2}: \quad P_{\Omega}(X)=P_{\Omega}\left(X^{\star}\right)\right. \\
& \left.\quad e_{S_1}^T X=0\quad \forall S_1 \in \mathcal{H}_1, X e_{S_2}=0\quad \forall S_2 \in \mathcal{H}_2\right\} \neq \emptyset
\end{aligned}
\end{equation}
If, in addition, \eqref{consis} consists of a single point we say that $\mathcal{H}_1$
and $\mathcal{H}_2$ are uniquely consistent with $P_{\Omega}\left(X^{\star}\right)$.

It was showed that if $\mathcal{C} \subseteq \mathbb{F}_2^n$ is a subspace of dimension $0 \leq r \leq$ $n-1$. Then $\mathcal{C}^{\perp}=\left[\mathcal{C}^{\perp}\right]_{\leq r+1}$

 Therefore, if we set $\left|S_1\right|=\left|S_2\right|\leq r+1$, and if both of $\operatorname{span}\left\{e_{S_1}: \mathcal{H}_{1, r+1}\right\}=\left[\mathcal{C}_{\text {col }}^{\perp}\right]_{\leq r+1}=\mathcal{C}^{\perp}_{\text {col }}$ and $\operatorname{span}\left\{e_{S_2}: \mathcal{H}_{2, r+1}\right\}=\left[\mathcal{C}_{\text {row }}^{\perp}\right]_{\leq r+1}=\mathcal{C}^{\perp}_{\text {row }}$ occur, then $\mathcal{H}_{1, r+1}$ and $\mathcal{H}_{2, r+1}$ are consistent with $P_{\Omega}\left(X^{\star}\right)$. And if $\mathcal{H}_{1, r+1}$ and $\mathcal{H}_{2, r+1}$ are uniquely consistent with $P_{\Omega}\left(X^{\star}\right)$, then they can recover the matrix successfully with high probability.
 
 The way to construct $\mathcal{H}_{1, s_1}$ and $\mathcal{H}_{1, s_2}$ is the following algorithm. 
 \begin{algorithm}[h]
 \caption{Constructing $\mathcal{H}_{1, s_1}$ }
 \text { Input: } $\Omega \subseteq\left[n_1\right] \times\left[n_2\right], P_{\Omega}\left(X^{\star}\right)$ \text {, positive integer }$ s_1$ \\
 \text { 1: } $\mathcal{H}_{1, s_1} \leftarrow \emptyset$ \\
 \text { 2: for } $S_1 \subseteq\left[n_1\right],\left|S_1\right| \leq s_1$ \text { do } \\
 \text { 3: } $\quad T_1 \leftarrow \bigcap_{i \in S_1}\left\{j \in\left[n_2\right]:(i, j) \in \Omega\right\} $\\
 \text { 4: }$ \quad$ \text { if }$ \sum_{i \in S_1} \bX_{i j}=0$ \text { for all }$ j \in T_1$ \text { then } \\
 \text { 5: } $\quad \quad \mathcal{H}_{1, s_1} \leftarrow \mathcal{H}_{1, s_1} \cup \left\{S_1\right\} $\\
 \text { 6: } $\quad $\text { end if } \\
 \text { 7: end for }
\end{algorithm}

\textbf{Step1}: Prove that $\mathcal{H}_{1, r+1}$ and $\mathcal{H}_{2, r+1}$ are consistent with $P_{\Omega}\left(X^{\star}\right)$ with probability at least $1-2\theta$.Where we set $s_1=s_2=r+1$.
 
If we fix $S_1 \subseteq\left[n_1\right]$ then  $T_1 \sim$ $\mathcal{B}\left(\left[n_2\right], p^{\left|S_1\right|}\right)$ since any $j \in T_1$ if and only if $(i, j) \in \Omega$ for all $i \in S_1$. Let $x=e_{S_1}^T X^{\star} \in \mathcal{C}_{\text {row }}$. Suppose $P_{T_1}(x)=0$, or equivalently that $S_1 \in \mathcal{H}_{1, r+1}$. Then the probability that $e_{S_1} \notin\left[\mathcal{C}_{\text {col }}^{\perp}\right]_{\leq r+1}$ (i.e. $x \neq 0$ ) is bounded above by $P_e(\epsilon=1-p^{\left|S_1\right|}, n,\dim \cC)$.

Taking a union bound over all $S_1 \subseteq\left[n_1\right]$ with $\left|S_1\right| \leq r+1$, the probability that $\operatorname{span}\left\{e_{S_1}: S_1 \in \mathcal{H}_{1, r+1}\right\} \neq\left[\mathcal{C}_{\text {col }}^{\perp}\right]_{\leq r+1}$ is at most

\begin{equation}
\begin{aligned}
&\sum_{k=1}^{r+1}\left(\begin{array}{c}
n \\
k
\end{array}\right) P_e(\epsilon=1-p^{r+1}, n,\dim \cC)
\\ 
&\leq n^{r+1} P_e(\epsilon=1-p^{r+1}, n,\dim \cC) \leq \theta 
\end{aligned}
\end{equation}

Similarly, span $\left\{e_{S_2}: S_2 \in \mathcal{H}_{2, r+1}\right\} \neq\left[\mathcal{C}_{\text {row }}^{\perp}\right]_{\leq r+1}$ with probability at most $\theta$. Take a union bound with this two events, we can get that for  $p>\tilde{p}$, where $\tilde{p}>p'$,$\mathcal{H}_{1, r+1}$ and $\mathcal{H}_{2, r+1}$ are consistent with $P_{\Omega}\left(X^{\star}\right)$ with probability more than $1-2\theta$. The parameters $p'$ and $\tilde{p}$ are the solutions to 

\be{eq}
n^{r+1}P_e(1-p^{r+1},n,r)=\theta,
\ee

\textbf{Step2} Prove that $\mathcal{H}_{1, r+1}$ and $\mathcal{H}_{2, r+1}$ are uniquely consistent with $P_{\Omega}\left(X^{\star}\right)$ with probability at least $1-3\theta$.

It was showed that
$\left[\mathcal{C}_{\text {col }}^{\perp}\right]_{\leq r+1}=\mathcal{C}^{\perp}_{\text {col }}$,
$\left[\mathcal{C}_{\text {row }}^{\perp}\right]_{\leq r+1}=\mathcal{C}^{\perp}_{\text {row }}$.
With Corollary 1 in \cite{saunderson2016simple}, we get that the $p^{\star}>p_0$ and $p'>p_0$, where
$$
p_0=\frac{\operatorname{dim}\left(\mathcal{C}_{\text{col}}\right) \operatorname{dim}\left(\mathcal{C}_{\text{row}}\right) \log (2)+\log (1 / \epsilon)}{d\left(\mathcal{C}_{\text{col}}\right) d\left(\mathcal{C}_{\text{row}}\right)}
$$
With Theorem 1 in \cite{saunderson2016simple},  $\mathcal{H}_{1, r+1}$ and $\mathcal{H}_{2, r+1}$ are  uniquely consistent with $P_{\Omega}\left(X^{\star}\right)$ with probability at least $1-3\theta$. That means we can recover $X^{\star}$ with $\mathcal{H}_{1, r+1}$ and $\mathcal{H}_{2, r+1}$ with probability at least $1-3\theta$.

\section{Additional Plot}\label{error_plot}
The plot below shows the gap between the result of Lemma\,1 in \cite{saunderson2016simple} and \eqref{ashikhmin}.

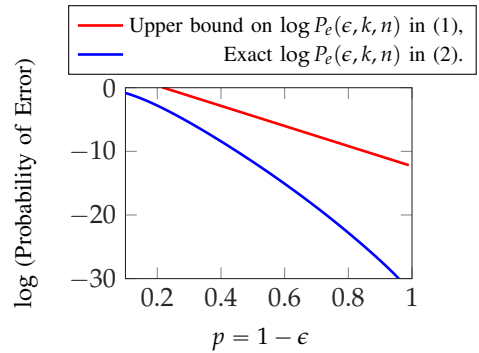
\begin{figure}[h]
 	 \begin{minipage}{.5\textwidth}
		\centering
%

\begin{tikzpicture}

\definecolor{mycolor1}{rgb}{0.15,0.15,0.15}
\definecolor{mycolor2}{rgb}{0,0,1}
\definecolor{mycolor3}{rgb}{1,0,0}
\definecolor{mycolor4}{rgb}{0,0.45,0.74}
\definecolor{mycolor5}{rgb}{0.64,0.08,0.18}
\definecolor{mycolor6}{rgb}{0.2,0.6,0.2}
\definecolor{mycolor7}{rgb}{1,0 ,1}

\begin{axis}[
x label style={at={(axis description cs:0.47,0)},anchor=north},
y label style={at={(axis description cs:0.05,0.5)},rotate=0,anchor=south},
xlabel={\small $p=1-\epsilon$},
ylabel={\small $\log$ (Probability of Error)},
scale only axis,
every outer x axis line/.append style={mycolor1},
every x tick label/.append style={font=\color{mycolor1}},
every outer y axis line/.append style={mycolor1},
every y tick label/.append style={font=\color{mycolor1}},
width=1.5in,
height=1in,
xmin=0.1, xmax=1,
ymin=-30, ymax=0,
axis on top,
legend entries={{ \small Upper bound on $\log P_e(\epsilon,k,n)$ in \eqref{Saunderson_bound},}, { \small Exact $\log P_e(\epsilon,k,n)$ in \eqref{ashikhmin}. },},
legend style={ nodes={scale=.9, transform shape}, legend columns=1,at={(-0.2,1.25)},anchor=west},
legend cell align=right,
mark options={solid,scale=1.3}
]
\addplot [
color=red,
solid,
line width=1.0pt,
]
coordinates{
	(0.215,0)
	(0.22,-0.0104769)
	(0.225,-0.0894817)
	(0.23,-0.168487)
	(0.235,-0.247491)
	(0.24,-0.326496)
	(0.245,-0.405501)
	(0.25,-0.484506)
	(0.255,-0.563511)
	(0.26,-0.642516)
	(0.265,-0.72152)
	(0.27,-0.800525)
	(0.275,-0.87953)
	(0.28,-0.958535)
	(0.285,-1.03754)
	(0.29,-1.11654)
	(0.295,-1.19555)
	(0.3,-1.27455)
	(0.305,-1.35356)
	(0.31,-1.43256)
	(0.315,-1.51157)
	(0.32,-1.59057)
	(0.325,-1.66958)
	(0.33,-1.74858)
	(0.335,-1.82759)
	(0.34,-1.90659)
	(0.345,-1.9856)
	(0.35,-2.0646)
	(0.355,-2.14361)
	(0.36,-2.22261)
	(0.365,-2.30162)
	(0.37,-2.38062)
	(0.375,-2.45963)
	(0.38,-2.53863)
	(0.385,-2.61764)
	(0.39,-2.69664)
	(0.395,-2.77565)
	(0.4,-2.85465)
	(0.405,-2.93366)
	(0.41,-3.01266)
	(0.415,-3.09167)
	(0.42,-3.17067)
	(0.425,-3.24968)
	(0.43,-3.32868)
	(0.435,-3.40768)
	(0.44,-3.48669)
	(0.445,-3.56569)
	(0.45,-3.6447)
	(0.455,-3.7237)
	(0.46,-3.80271)
	(0.465,-3.88171)
	(0.47,-3.96072)
	(0.475,-4.03972)
	(0.48,-4.11873)
	(0.485,-4.19773)
	(0.49,-4.27674)
	(0.495,-4.35574)
	(0.5,-4.43475)
	(0.505,-4.51375)
	(0.51,-4.59276)
	(0.515,-4.67176)
	(0.52,-4.75077)
	(0.525,-4.82977)
	(0.53,-4.90878)
	(0.535,-4.98778)
	(0.54,-5.06679)
	(0.545,-5.14579)
	(0.55,-5.2248)
	(0.555,-5.3038)
	(0.56,-5.38281)
	(0.565,-5.46181)
	(0.57,-5.54082)
	(0.575,-5.61982)
	(0.58,-5.69883)
	(0.585,-5.77783)
	(0.59,-5.85683)
	(0.595,-5.93584)
	(0.6,-6.01484)
	(0.605,-6.09385)
	(0.61,-6.17285)
	(0.615,-6.25186)
	(0.62,-6.33086)
	(0.625,-6.40987)
	(0.63,-6.48887)
	(0.635,-6.56788)
	(0.64,-6.64688)
	(0.645,-6.72589)
	(0.65,-6.80489)
	(0.655,-6.8839)
	(0.66,-6.9629)
	(0.665,-7.04191)
	(0.67,-7.12091)
	(0.675,-7.19992)
	(0.68,-7.27892)
	(0.685,-7.35793)
	(0.69,-7.43693)
	(0.695,-7.51594)
	(0.7,-7.59494)
	(0.705,-7.67395)
	(0.71,-7.75295)
	(0.715,-7.83196)
	(0.72,-7.91096)
	(0.725,-7.98997)
	(0.73,-8.06897)
	(0.735,-8.14798)
	(0.74,-8.22698)
	(0.745,-8.30598)
	(0.75,-8.38499)
	(0.755,-8.46399)
	(0.76,-8.543)
	(0.765,-8.622)
	(0.77,-8.70101)
	(0.775,-8.78001)
	(0.78,-8.85902)
	(0.785,-8.93802)
	(0.79,-9.01703)
	(0.795,-9.09603)
	(0.8,-9.17504)
	(0.805,-9.25404)
	(0.81,-9.33305)
	(0.815,-9.41205)
	(0.82,-9.49106)
	(0.825,-9.57006)
	(0.83,-9.64907)
	(0.835,-9.72807)
	(0.84,-9.80708)
	(0.845,-9.88608)
	(0.85,-9.96509)
	(0.855,-10.0441)
	(0.86,-10.1231)
	(0.865,-10.2021)
	(0.87,-10.2811)
	(0.875,-10.3601)
	(0.88,-10.4391)
	(0.885,-10.5181)
	(0.89,-10.5971)
	(0.895,-10.6761)
	(0.9,-10.7551)
	(0.905,-10.8341)
	(0.91,-10.9131)
	(0.915,-10.9921)
	(0.92,-11.0712)
	(0.925,-11.1502)
	(0.93,-11.2292)
	(0.935,-11.3082)
	(0.94,-11.3872)
	(0.945,-11.4662)
	(0.95,-11.5452)
	(0.955,-11.6242)
	(0.96,-11.7032)
	(0.965,-11.7822)
	(0.97,-11.8612)
	(0.975,-11.9402)
	(0.98,-12.0192)
	(0.985,-12.0982)
	(0.99,-12.1772)
	
};

\addplot [
color=blue,
solid,
line width=1.0pt,
]
coordinates{
	(0.1,-0.835104)
	(0.105,-0.909029)
	(0.11,-0.986351)
	(0.115,-1.06689)
	(0.12,-1.15049)
	(0.125,-1.23701)
	(0.13,-1.32631)
	(0.135,-1.41826)
	(0.14,-1.51277)
	(0.145,-1.60971)
	(0.15,-1.709)
	(0.155,-1.81053)
	(0.16,-1.91423)
	(0.165,-2.02)
	(0.17,-2.12777)
	(0.175,-2.23748)
	(0.18,-2.34904)
	(0.185,-2.46238)
	(0.19,-2.57746)
	(0.195,-2.69419)
	(0.2,-2.81254)
	(0.205,-2.93243)
	(0.21,-3.05382)
	(0.215,-3.17666)
	(0.22,-3.30089)
	(0.225,-3.42647)
	(0.23,-3.55336)
	(0.235,-3.68151)
	(0.24,-3.81089)
	(0.245,-3.94144)
	(0.25,-4.07314)
	(0.255,-4.20595)
	(0.26,-4.33983)
	(0.265,-4.47475)
	(0.27,-4.61068)
	(0.275,-4.74759)
	(0.28,-4.88545)
	(0.285,-5.02424)
	(0.29,-5.16392)
	(0.295,-5.30447)
	(0.3,-5.44587)
	(0.305,-5.5881)
	(0.31,-5.73113)
	(0.315,-5.87495)
	(0.32,-6.01953)
	(0.325,-6.16486)
	(0.33,-6.31092)
	(0.335,-6.45769)
	(0.34,-6.60517)
	(0.345,-6.75332)
	(0.35,-6.90216)
	(0.355,-7.05165)
	(0.36,-7.20179)
	(0.365,-7.35257)
	(0.37,-7.50397)
	(0.375,-7.656)
	(0.38,-7.80864)
	(0.385,-7.96188)
	(0.39,-8.11571)
	(0.395,-8.27014)
	(0.4,-8.42515)
	(0.405,-8.58075)
	(0.41,-8.73691)
	(0.415,-8.89365)
	(0.42,-9.05096)
	(0.425,-9.20884)
	(0.43,-9.36727)
	(0.435,-9.52627)
	(0.44,-9.68583)
	(0.445,-9.84594)
	(0.45,-10.0066)
	(0.455,-10.1678)
	(0.46,-10.3296)
	(0.465,-10.492)
	(0.47,-10.6549)
	(0.475,-10.8183)
	(0.48,-10.9824)
	(0.485,-11.1469)
	(0.49,-11.3121)
	(0.495,-11.4778)
	(0.5,-11.6441)
	(0.505,-11.8109)
	(0.51,-11.9783)
	(0.515,-12.1463)
	(0.52,-12.3149)
	(0.525,-12.484)
	(0.53,-12.6538)
	(0.535,-12.8241)
	(0.54,-12.995)
	(0.545,-13.1665)
	(0.55,-13.3385)
	(0.555,-13.5112)
	(0.56,-13.6845)
	(0.565,-13.8584)
	(0.57,-14.0329)
	(0.575,-14.208)
	(0.58,-14.3838)
	(0.585,-14.5601)
	(0.59,-14.7371)
	(0.595,-14.9147)
	(0.6,-15.093)
	(0.605,-15.2718)
	(0.61,-15.4514)
	(0.615,-15.6315)
	(0.62,-15.8124)
	(0.625,-15.9939)
	(0.63,-16.176)
	(0.635,-16.3588)
	(0.64,-16.5423)
	(0.645,-16.7265)
	(0.65,-16.9113)
	(0.655,-17.0968)
	(0.66,-17.283)
	(0.665,-17.47)
	(0.67,-17.6576)
	(0.675,-17.8459)
	(0.68,-18.0349)
	(0.685,-18.2247)
	(0.69,-18.4152)
	(0.695,-18.6064)
	(0.7,-18.7983)
	(0.705,-18.991)
	(0.71,-19.1844)
	(0.715,-19.3786)
	(0.72,-19.5735)
	(0.725,-19.7692)
	(0.73,-19.9657)
	(0.735,-20.1629)
	(0.74,-20.3609)
	(0.745,-20.5598)
	(0.75,-20.7594)
	(0.755,-20.9598)
	(0.76,-21.161)
	(0.765,-21.363)
	(0.77,-21.5658)
	(0.775,-21.7695)
	(0.78,-21.974)
	(0.785,-22.1794)
	(0.79,-22.3856)
	(0.795,-22.5926)
	(0.8,-22.8006)
	(0.805,-23.0093)
	(0.81,-23.219)
	(0.815,-23.4296)
	(0.82,-23.6411)
	(0.825,-23.8534)
	(0.83,-24.0667)
	(0.835,-24.2809)
	(0.84,-24.4961)
	(0.845,-24.7122)
	(0.85,-24.9293)
	(0.855,-25.1474)
	(0.86,-25.3665)
	(0.865,-25.5867)
	(0.87,-25.8079)
	(0.875,-26.0301)
	(0.88,-26.2536)
	(0.885,-26.4781)
	(0.89,-26.704)
	(0.895,-26.9311)
	(0.9,-27.1596)
	(0.905,-27.3896)
	(0.91,-27.6212)
	(0.915,-27.8546)
	(0.92,-28.09)
	(0.925,-28.3277)
	(0.93,-28.5681)
	(0.935,-28.8116)
	(0.94,-29.0589)
	(0.945,-29.3107)
	(0.95,-29.5683)
	(0.955,-29.833)
	(0.96,-30.1071)
	(0.965,-30.3937)
	(0.97,-30.6975)
	(0.975,-31.0257)
	(0.98,-31.3909)
	(0.985,-31.8165)
	(0.99,-32.3556)
	
};

\end{axis}

\end{tikzpicture} 
		\caption{ \small Comparison between the upper bound on $\log P_e(\epsilon,k,n)$ based on  \eqref{Saunderson_bound} and $\log P_e(\epsilon,k,n)$, characterized in \eqref{ashikhmin} for $k=5$ and $n=50$.}\label{logpe_comp}
	\end{minipage}
\end{figure}

\end{document}